# Structural and Magnetic Characterization of Spin Canted Mixed Ferrite-Cobaltites: LnFe$_{0.5}$Co$_{0.5}$O$_3$ (Ln = Eu and Dy)

Ashish Shukla[1], Oleg I. Lebedev[2], Md. Motin Seikh[3] and Asish K. Kundu[1]*

[1]*Discipline of Physics, Indian Institute of Information Technology, Design & Manufacturing Jabalpur, Dumna Airport Road, Madhya Pradesh–482005, India*

[2]*Laboratoire CRISMAT, ENSICAEN UMR6508, 6 Bd Maréchal Juin, Cedex 4, Caen-14050, France*

[3]*Department of Chemistry, Visva-Bharati University, Santiniketan, West Bengal –731235, India*

**Abstract**

The mixed ferrite-cobaltites LnFe$_{0.5}$Co$_{0.5}$O$_3$, with Ln = Eu & Dy have been prepared by a sol-gel method and the samples have been characterized using X-ray diffraction and electron microscopy. The magnetic investigations reveal that both samples ordered in canted antiferromagnetic structures near room temperature. The Dzyaloshinskii-Moriya or antisymmetric exchange interaction induces weak ferromagnetism due to canting of the antiferromagnetically ordered spins. In the case of Ln-Fe-Co orthoferrites, two magnetic sublattices (Ln$^{3+}$-4f and Fe$^{3+}$/Co$^{3+}$-3d) generally align in opposite directions and interesting temperature dependent phenomena: e.g. uncompensated antiferromagnetic sublattices and spin-reorientations, are observed in the system. The existence of hysteresis at low temperature region has been explained in terms of the strength of magnetic interactions between Fe$^{3+}$ and Co$^{3+}$ ions with different A-site rare earth cations.

# 1. Introduction

Orthoferrite perovskites LnFeO$_3$ have attracted researchers for decades due to their interesting magnetic, electrical and gas sensing properties [1-5]. Among orthoferrites, particular interest is focused on doped phases because this structural framework can accommodate a wide variety of cations on the 12-coordinated A-site and 6-coordinated B-site. Such structural flexibility provides large freedom to prepare new phases and/or modification of existing materials with new properties. These captivating physical properties have encouraged researchers to design various electrical and chemical sensors [4,5], catalysts [6,7], solid oxide fuel cells [8,9] and numerous other devices based on utility. Moreover, the magnetic interactions between lanthanide (Ln$^{3+}$-4f) and transition metal (Fe$^{3+}$-3d) in these perovskites are still an open research topic since they greatly influence the multiferroic behavior [2,10-13]. However, the 3d–4f coupling is difficult to investigate experimentally since it is generally much weaker compared to 3d–3d interaction in most perovskite structures [14]. In order to realize the 3d-4f coupling at the cost of the 3d-3d coupling in orthoferrites, the iron cation has been partially replaced by other magnetic cation on the B sites. Therefore, the B-site doping effect in LnFe$_{0.5}$Co$_{0.5}$O$_3$ orthoferrites have been recently investigated for their interesting physical properties, and these materials have also been found to be promising materials for multiferroics [12,13,15,16]. The first member of the series i.e. LaFe$_{0.5}$Co$_{0.5}$O$_3$ prepared by soft chemical synthesis, crystallizes in orthorhombic structure (Pbnm space group) and it is a canted antiferromagnet with T$_N$~ 370 K [15], which is higher than the solid state synthesized sample [16]. Therefore, we adopted the similar route of synthesis and reported a unique nano-scale ordered layered phase for SmFe$_{0.5}$Co$_{0.5}$O$_3$ [13], which also exhibits antiferromagnetism around room temperature (T$_N$~ 310 K). Nevertheless, latter perovskite depicts intrinsic magneto-dielectric coupling at low temperature. Recently, Lohr et al [12] have reported the synthesis and physical properties for LnFe$_{0.5}$Co$_{0.5}$O$_3$ perovskites with Ln = Tb, Dy, Ho, Er and Tm. These compounds exhibit antiferromagnetic transition ($T_N$) around the temperature range of 254-249K and there is an appearance of electric polarization in the lower size lanthanide ($TmFe_{0.5}Co_{0.5}O_3$). It is worth mentioning that with decreasing the lanthanide cation $T_N$ value decreases, which also supported by the data from our group [13,15]. Nevertheless, the transition temperature $T_N$ is almost 50 K lower for $LnFe_{0.5}Co_{0.5}O_3$ series and also the temperature variation with size is significantly weaker compared to La and Sm-phases [13,15]. Therefore, we considered it important to investigate the missing member i.e. $EuFe_{0.5}Co_{0.5}O_3$ in the present series along with the reported $DyFe_{0.5}Co_{0.5}O_3$ compound, which shows higher $T_N$ (~300 K) in the present investigation. Despite the interest in doped orthoferrites, their synthesis in pure phase remains a challenge, principally because high synthesis temperatures generally required for preparing perovskite orthoferrites. Also, entropy plays a dominant role in determining the pure phase and stable structure of reaction products. Previous investigations on the doped orthoferrites have motivated us to synthesize doped perovskite with smaller size of lanthanides in similar conditions [13,15]. Therefore, we have utilized sol-gel route synthesis to avoid higher temperature as it has explicit inherent advantages over the solid-state reaction synthesis methods. This controls the reaction pathways on a molecular level during the transformation of the precursor materials to the final product. Both the $EuFe_{0.5}Co_{0.5}O_3$ and $DyFe_{0.5}Co_{0.5}O_3$ perovskites were prepared below 1000 °C and detailed variable temperature magnetic measurements were performed. The results of these investigations include the unusual and complex magnetic behavior of $DyFe_{0.5}Co_{0.5}O_3$ with a $T_N$ of 300 K, which is much higher than the reported value i.e. 252 K [12]. Likewise, orthoferrite $EuFe_{0.5}Co_{0.5}O_3$ also exhibits antiferromagnetism around room temperature. In the case of orthoferrites the two magnetic (Ln and Fe-Co) sublattices generally ordered magnetically at different temperatures and show interesting phenomena, which have been investigated in details.

## 2. Experimental procedure

Soft chemical route method sol–gel was used for synthesizing $LnFe_{0.5}Co_{0.5}O_3$ (Ln = Eu and Dy) orthoferrites. The metal nitrates used in the synthesis were dissolved in distilled water along with citric acid (molar ration 1:2). The resulting solutions were heated for 3 hours at 60 °C and evaporated at 100 °C for gel formation. The gel mixture was dried at 150 °C for 12 h. Thereafter the resulting powders were allowed to decompose in air at 250 °C for 24 h. The resulting powders sintered at several temperatures ranging from 400 °C to 800 °C with intermediate grinding. The powder samples were again ground thoroughly and pressed into pellets, and finally sintered for 48 h in air at 927 °C (for $EuFe_{0.5}Co_{0.5}O_3$) and 950 °C (for $DyFe_{0.5}Co_{0.5}O_3$) respectively. The sintered pellets were used for various measurements and ground again to form fine powder for recording XRD patterns in the 2θ range of 10°-100° with a step size of 0.02°. The powder X-ray diffraction (PXRD) data were recorded with a Philips X-Pert diffractometer employing Cu-K$_\alpha$ radiation (λ = 1.5418Å). The phases were indexed by performing *Rietveld* method using the FullProf program [17]. The transmission electron microscopy (TEM) including electron diffraction (ED) and high resolution TEM (HRTEM) experiments were performed using aberration double-corrected JEM ARM 200F microscope operated at 200 kV equipped with CENTURIO EDX detector. TEM samples were prepared by dropping the colloidal solutions of powders in methanol onto a holey Cu carbon grid followed by the evaporation of the solvent. The field and temperature dependence of magnetization under various conditions was investigated using a physical property measurement system (PPMS, Quantum Design) and the system was cleaned after each measurement for remanent field.

## 3. Results and discussion

The powder X-ray diffraction (PXRD) patterns of $EuFe_{0.5}Co_{0.5}O_3$ and $DyFe_{0.5}Co_{0.5}O_3$ could be indexed with the orthorhombic symmetry in the *Pnma* space group, as concluded from Rietveld refinement. The refined structural parameters of samples based on XRD patterns are shown in Fig. 1 and the corresponding lattice parameters are presented in Table-1. The difference in the values of lattice parameters of two samples could be attributed to the size differences between the rare earth cations.

From PXRD study, we obtained information about the crystal structure however the distribution of different phases, crystallinity of grains etc., cannot be determined from it. In order to determine those factors, Transmission Electron Microscopy (TEM) investigations have been carried out. The electron diffraction (ED) patterns (Fig. 2 & Fig. 3a) confirm the crystalline nature and orthorhombic *Pnma* structures of both the samples. Both mixed ferrite-cobaltites show similar size of grains, which are clearly visible in the grain size distribution at different scale (Fig. 2 & Fig. 3b). High angle annular dark field scanning TEM (HAADF-STEM) was acquired in parallel with the energy-dispersive X-ray spectroscopy (EDX) mapping (Fig. 3c) shows that the distribution of the Dy, Fe and Co elements is uniform with an atomic ratio close to nominal composition. High resolution TEM (HRTEM) imaging (Fig 3d) confirms good crystallinity of the grains, which are free from any defects.

Fig. 4 shows the temperature dependent magnetization, M(T), for $EuFe_{0.5}Co_{0.5}O_3$ in the zero field cooled (ZFC), field cooled cooling (FCC) and field cooled warming (FCW) conditions in the fields of 50 Oe and 1000 Oe. The ZFC-FCW (FCC) data show large thermomagnetic irreversibility below 300 K with a clear signature of magnetic transition ($T_N \sim 300$ K) and finally they merge above room temperature. The ZFC data exhibits a peak around the transition point (~300 K). The FC-magnetization value increases gradually with decreasing temperature and depicts a sharp transition at $T_N$. The maximum value of magnetic moment is merely 0.019 $\mu_B$/f.u. (Fig. 4b), signifying an antiferromagnetic (AFM) type interaction between the $Fe^{3+}$ and $Co^{3+}$ cations similar to the reported $LnFe_{0.5}Co_{0.5}O_3$ mixed ferrite-cobaltites [12].

In order to correlate the effect of $Dy^{3+}$ cation (considering the contribution of 4f electrons) on magnetism, we have also carried out magnetization, M(T), investigation for $DyFe_{0.5}Co_{0.5}O_3$ mixed ferrite-cobaltites. Fig. 5 shows the temperature dependent ZFC, FCC and FCW magnetization for $DyFe_{0.5}Co_{0.5}O_3$ in the applied fields of H = 50 Oe and 1000 Oe, the curves exhibit a broad magnetic transition near the room temperature akin to $EuFe_{0.5}Co_{0.5}O_3$ perovskite. The M(T) data for $DyFe_{0.5}Co_{0.5}O_3$ perovskite exhibits various magnetic transitions down to lowest temperature in the low applied field (Fig.5a) and corroborate the behavior of higher field as reported earlier [12]. The 50 Oe FCC/FCW magnetization value increases rapidly below $T_N$ (~ 300 K) and a sharp drop in magnetization appears around 90 K. Further decreasing temperature (< 70 K) the moment increases and shows broad valley around 60 K. We have also noticed the thermal hysteresis below $T_N$ between the FCC and FCW data (Fig. 5a) with some unusual behavior as far as the thermal hysteresis is concerned. They exhibit higher value of magnetization for FCC compared to FCW data in the thermal hysteresis region (60-300 K), which could give rise to metastability in the system. The higher field ZFC and FCC/FCW data (Fig. 5b) show weak thermal hysteresis and follow similar behavior in the 10-400 K temperature range.

The complicated magnetic behavior of both mixed ferrite-cobaltites in measured temperature range could arise due to the two inequivalent magnetic ordering at low temperature between the cations in different magnetic sublattices, i.e. 4f electron based Ln sublattice and 3d electron based Fe-Co sublattice, which are arranged anti-parallelly [11,12]. Moreover, the AFM ordering of Fe-Co sublattice conforms with the AFM behaviour of $Fe^{3+}$-

O-$Co^{3+}$, $Fe^{3+}$-O-$Fe^{3+}$ and $Co^{3+}$-O-$Co^{3+}$ interactions [12,13,15]. The magnetic transition near 300 K for both perovskites may be due to canting of the antiferromagnetically aligned magnetic spins (which is well known for orthoferrites). This leads to weak ferromagnetic state related to non-collinear antiferromagnetism in the Fe sublattice similar to Dzyaloshinskii-Moriya interaction [2,18]. However, Ln-sublattice order antiferromagnetically at relatively low temperatures [11,12]. Here, it is important to mention that the $T_N$ for present $DyFe_{0.5}Co_{0.5}O_3$ compound is much higher than the reported value of 252 K [12]. This is because the synthesis methods play a vital role in determining the crystal structure and physical properties of the same compound, which we have reported earlier for mixed ferrite-cobaltites $LaFe_{0.5}Co_{0.5}O_3$ [15].

For understanding the exact behaviour of magnetic interactions below room temperature, the field dependent isothermal magnetization needs to be investigated at different temperatures. Fig. 6 exhibits the field dependent magnetization, M(H), at 10 K and 100 K for $EuFe_{0.5}Co_{0.5}O_3$. The sample depicts hysteresis loop with low remanent magnetization (0.01 $\mu_B$/f.u.) and large coercivity (4.2 kOe) at 10 K. Importantly, the saturation in the magnetization value is not attained in the applied field and the graph indicates almost linear behaviour with increasing field. The highest obtained magnetization value is only 0.13 $\mu_B$/f.u. for H = 50 kOe. Nevertheless, the M(H) behavior at 100 K exhibits higher values of remnant (0.013 $\mu_B$/f.u.) and coercivity (6.5 kOe) compared to 10 K. This may be the consequence of large domain wall pinning of magnetic spins at 100 K. Thus, the magnetic hysteresis loops indicate that $EuFe_{0.5}Co_{0.5}O_3$ is a canted antiferromagnet in which the canting of magnetic spins results in uncompensated magnetic moment, leading to a weak ferromagnetic behavior similar to reported $LnFe_{0.5}Co_{0.5}O_3$ mixed ferrite-cobaltites [12,13,15]. We have also compared the isothermal magnetization, M(H), for $DyFe_{0.5}Co_{0.5}O_3$ perovskite under three different temperatures (Fig. 7) to correlate its AFM behaviour with $EuFe_{0.5}Co_{0.5}O_3$ perovskite. In contrast to latter perovskite, the M(H) curves exemplify weak hysteresis at 100 K ($M_r$~ 0.011 $\mu_B$/f.u. and $H_c$ ~ 520 Oe) and 80 K ($H_c$ ~ 320 Oe). However, there is no hysteresis at 10 K (inset Fig. 7a). The M(H) data show linear field dependency at higher field, without exhibiting magnetic saturation (the maximum magnetization value being 2 $\mu_B$/f.u. at 10 K). The weak hysteresis below $T_N$ suggests that the $DyFe_{0.5}Co_{0.5}O_3$ perovskite also manifests canted AFM type interactions among the cations similar to $EuFe_{0.5}Co_{0.5}O_3$ perovskite, which is clearly in accordance with the previously reported results [11-13,15,16]. Importantly, the $H_C$ values are higher for Eu-phase compared to Dy-phase, which could be due the AFM contribution of $Dy^{3+}$ cations at low temperature [18] (ground state of $Eu^{3+}$ to be non-magnetic[14]). Both the perovskites exhibit a weak ferromagnetic behavior below the room temperature, arising from a canting of the AFM spins. The DM interaction could be the basis of weak ferromagnetism in these mixed ferrite-cobaltites, which has been well established for perovskite orthoferrites [1-3,11-13].

## 4. Conclusions

In this work on Ln-Fe-Co-O systems (Ln = Eu & Dy), the sol-gel synthesis method enables the orthorhombic structure to be obtained as a pure phase. The magnetic properties investigation of these oxides show that both the compositions exhibit near room temperature magnetic transitions. The canted antiferromagnetic state is realized at low temperature for both the perovskites, which is most likely due to the ordering of 3d and 4f sublattices as reported in the literature for Ln-Fe-Co-O systems [12]. We hope our finding will motivate the researcher to obtain room temperature magneto-electric material, which is of current interest for future technological applications.

**Acknowledgements**

AKK and MMS thank the Science and Engineering Research Board (SERB), India for financial support through the project grant # EMR/2016/000083 and Prof. B. Raveau for his valuable comments and suggestions. We would also like to thank the reviewers for their constructive guidance during evaluation process.

**Table. 1.** Lattice parameters of LnFe$_{0.5}$Co$_{0.5}$O$_3$ (Ln = Eu, Dy) perovskites. Here *a*, *b*, *c* are the lattice parameters, R$_b$ and R$_f$ are the Bragg factor and fit factor, respectively.

| Perovskite | EuFe$_{0.5}$Co$_{0.5}$O$_3$ | DyFe$_{0.5}$Co$_{0.5}$O$_3$ |
|---|---|---|
| Space group | *Pnma* | *Pnma* |
| *a* (Å) | 5.488(4) | 5.505(6) |
| *b* (Å) | 7.577(5) | 7.504(6) |
| *c* (Å) | 5.315(3) | 5.239(4) |
| *V* (Å$^3$) | 221.074 | 216.421 |
| R$_b$(%) | 13.6 | 11.3 |
| R$_f$(%) | 32.7 | 19.9 |
| Chi-factor | 3.95 | 2.87 |

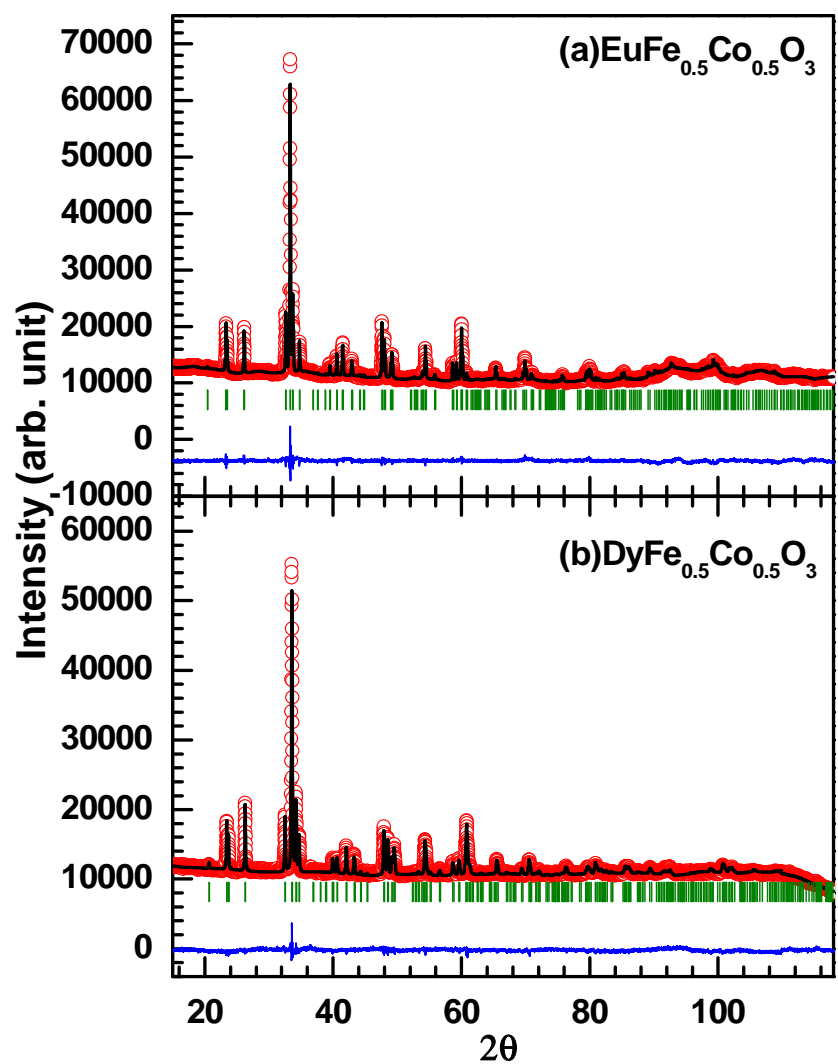

**Fig.1.** Rietveld analysed XRD pattern for (a) EuFe$_{0.5}$Co$_{0.5}$O$_3$ and (b) DyFe$_{0.5}$Co$_{0.5}$O$_3$. Open symbols are experimental data and the solid and vertical lines represent the difference curve and Bragg positions respectively.

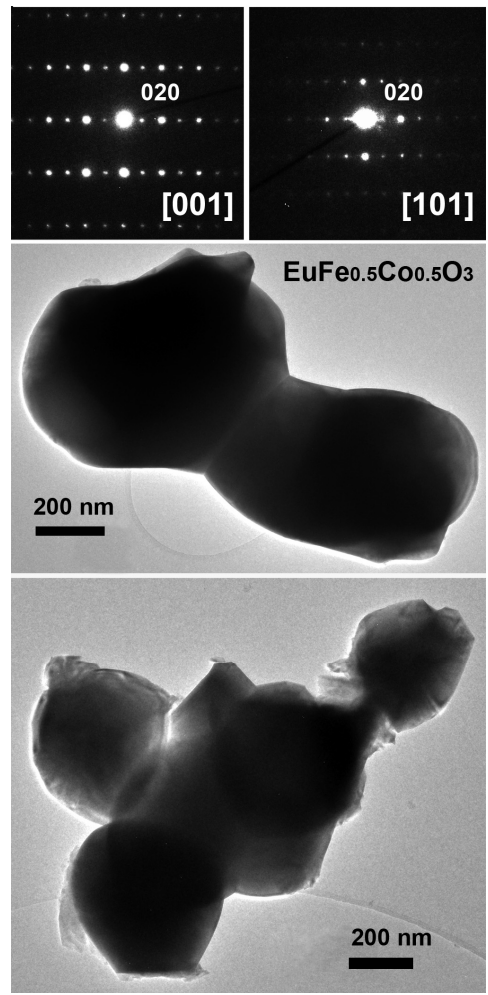

**Fig.2.** ED patterns along two main zone axis of EuFe$_{0.5}$Co$_{0.5}$O$_3$ sample indexed based on *Pnma* structure determine by PXRD (Table 1). Two representative low magnification bright field TEM images of grains shape and size distribution are presented in lower panel.

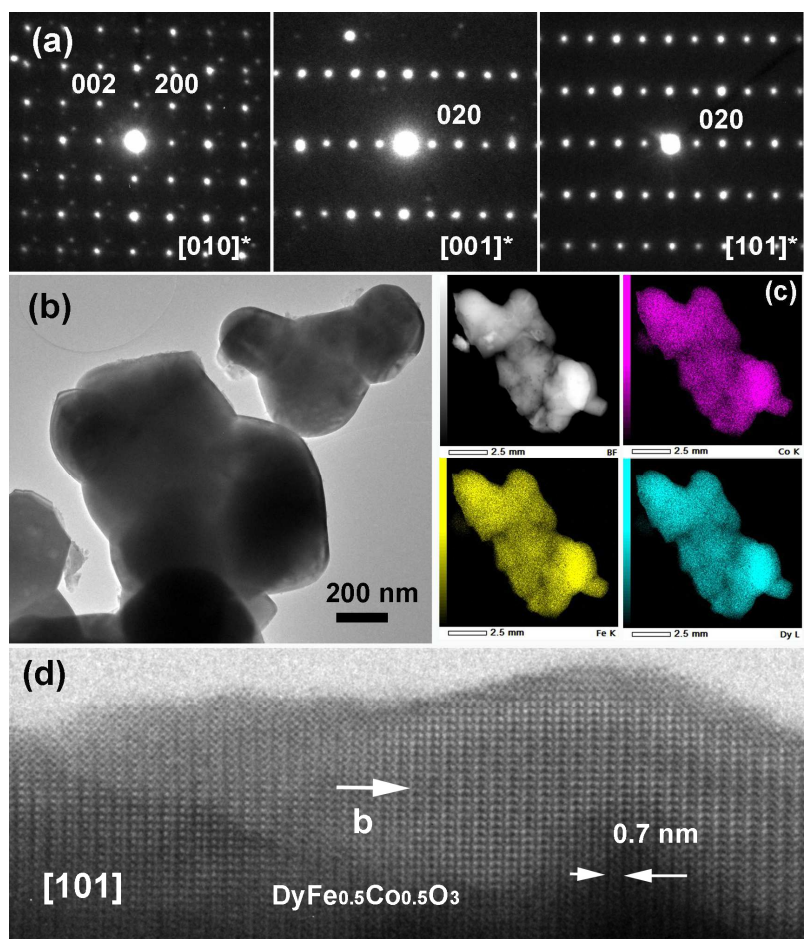

**Fig.3.** (a) ED patterns along main zone axis [010], [001] and [101]. All ED patterns indexed based on *Pnma* structure and parameters determined by PXRD (Table 1); (b) Low magnification bright field TEM image of $DyFe_{0.5}Co_{0.5}O_3$ sample; (c) HAADF-STEM image and parallel acquired EDX mapping of Co K (purple), Fe K (yellow) and Dy L (light blue) elements. (d) HRTEM image of $DyFe_{0.5}Co_{0.5}O_3$ grain viewed along [101] direction.

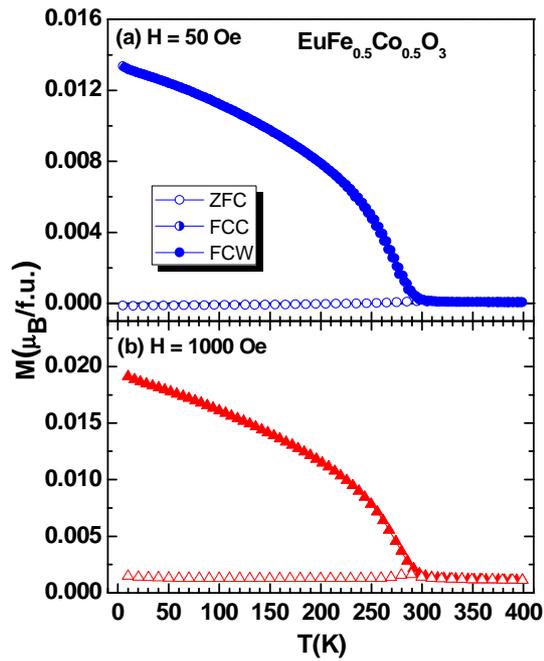

**Fig.4.** Temperature dependent ZFC (open symbol), FCC(half solid) and FCW (solid symbol) magnetization, M, plot for $EuFe_{0.5}Co_{0.5}O_3$ in the applied fields of (a) H=50 Oe and (b) H=1000 Oe.

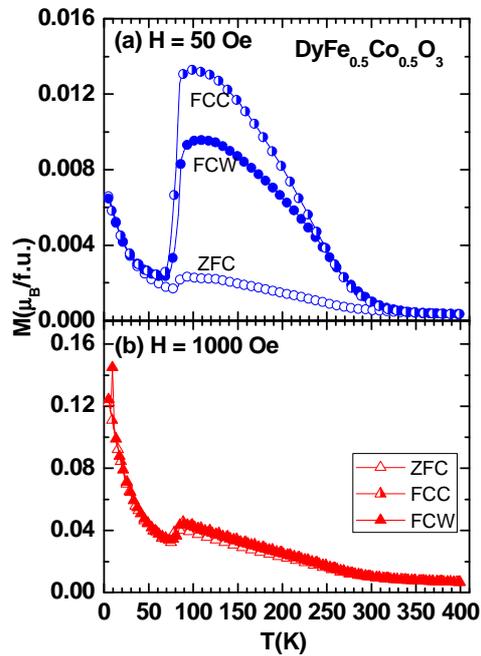

**Fig.5.** Temperature dependent ZFC (open symbol), FCC(half solid) and FCW (solid symbol) magnetization, M, plot for $DyFe_{0.5}Co_{0.5}O_3$ in the applied fields of (a) H=50 Oe and (b) H=1000 Oe.

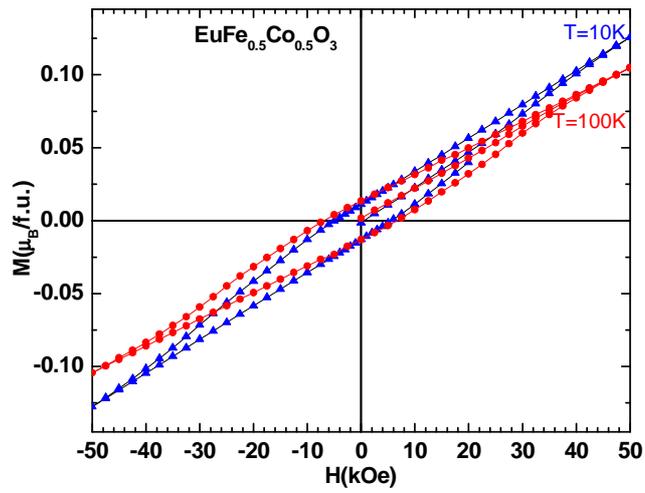

**Fig.6.** Field dependent isothermal magnetic hysteresis, M(H), loops at two different temperatures for $EuFe_{0.5}Co_{0.5}O_3$.

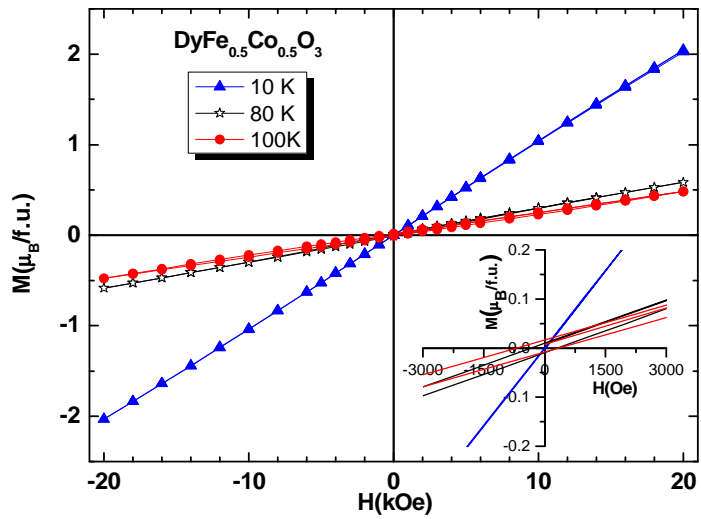

**Fig.7.** Field dependent isothermal magnetic hysteresis, M(H), loops at three different temperatures for $DyFe_{0.5}Co_{0.5}O_3$.